\input{aipcheck}

\documentclass[
    ,final            
  ]
  {aipproc}
\usepackage{amssymb}
\usepackage{amsmath}
\layoutstyle{6x9}

\begin{document}

\title{The Dihadron fragmentation functions way to Transversity }


\author{A.~Courtoy}{
  address={INFN-Pavia, Via Bassi 6, 27100 Pavia, Italy}
}

\author{Alessandro~Bacchetta}{
  address={Dipartimento di Fisica Nucleare e Teorica, Universit\`a di Pavia}
  ,altaddress={INFN-Pavia, Via Bassi 6, 27100 Pavia, Italy}
}

\author{Marco~Radici}{
  address={INFN-Pavia, Via Bassi 6, 27100 Pavia, Italy} 
}

\begin{abstract}
Observations of the transversity parton distribution 
based on an analysis of pion-pair production in deep
inelastic scattering off transversely polarized targets are presented. This extraction relies on the knowledge of dihadron fragmentation functions,
which are obtained from electron-positron annihilation measurements. This is the
first attempt to determine the transversity distribution in the framework of
collinear factorization.
\end{abstract}

\maketitle


The distribution of quarks and gluons inside hadrons can be described by means of
parton distribution functions (PDFs). In a parton-model picture, PDFs describe
combinations of number densities of quarks and gluons in a fast-moving
hadron. The knowledge of PDFs is crucial for our understanding of QCD and for the interpretation of high-energy 
experiments involving hadrons. 

In the Bjorken limit, the partonic structure of 
the nucleon is described in terms of only three collinear PDFs: the unpolarized, 
$f_1^q(x)$, and helicity, $g_1^q(x)$, distribution functions, and the transversity distribution 
function $h_1^q(x)$, whose extraction we present here. Transversity measures the transverse polarization of quarks with flavor $q$ 
and fractional momentum $x$ in a transversely polarized
nucleon.

There is no transversity for gluons in a nucleon, and $h_1^q$ has a pure non-singlet 
scale evolution~\cite{Artru:1990zv}. 
Transversity is a chiral-odd function and appears in cross sections combined with another chiral-odd function, e.g.,  the Collins fragmentation function  in the case of single-particle-inclusive DIS. The latter convolution  gives rise to a specific azimuthal
modulation of the cross section. The amplitude of the modulation has been
measured by the HERMES and COMPASS
collaborations~\cite{Airapetian:2010ds, Alekseev:2010rw}.
The first-ever extraction of $h_1^q$ arose from a simultaneous analysis of SIDIS data  and the
$e^+ e^- \to \pi \pi X$ data from Belle~\cite{Abe:2005zx}, from which the Collins
function can be determined~\cite{Anselmino:2007fs}. 

As for single-particle-inclusive DIS, there are two main issues: the convolution should be analyzed in the transverse-momentum-dependent factorization framework ; QCD evolution from Belle's scale to HERMES's  has to be applied coherently to that framework.

Here we extract transversity in an independent way requiring only standard collinear factorization, i.e. by considering the semi-inclusive deep-inelastic production of two  hadrons with small invariant mass, where the above complications are absent~\cite{Bacchetta:2011ip}. 
 That is, in this case, the transversity distribution function is multiplied by a
chiral-odd Dihadron Fragmentation Function (DiFF), denoted as
$H_1^{\sphericalangle\,q}$~\cite{Radici:2001na}, which describes the correlation between the 
transverse polarization of the fragmenting quark with flavor $q$ and the azimuthal 
orientation of the plane containing the momenta of the detected hadron pair. 
Contrary to the Collins mechanism, this effect survives after integration over quark transverse 
momenta and can be analyzed in the framework of collinear factorization.  This process has been 
studied from different perspectives in a number of
papers, e.g.~\cite{Efremov:1992pe}. 
The only published measurement of the relevant asymmetry has been presented by
the HERMES collaboration for the production of $\pi^+ \pi^-$ pairs on transversely polarized 
protons~\cite{Airapetian:2008sk}. Preliminary measurements have been presented
by the COMPASS collaboration~\cite{Wollny:2010}. 

Similarly to the single-hadron case, $H_1^{\sphericalangle\,q}$  has to be independently determined
by looking at correlations between the azimuthal orientations of two pion pairs in back-to-back 
jets in $e^+e^-$ annihilation~\cite{Artru:1996zu,Boer:2003ya,Bacchetta:2008wb}. The measurement of this so-called 
Artru--Collins azimuthal asymmetry has recently  become possible thanks to the Belle 
collaboration~\cite{Vossen:2011fk}.

The $A_{UT}^{\sin(\phi_{R} + \phi_S)\sin \theta}$ measured by HERMES~\cite{Airapetian:2008sk} can be interpreted 
as~\cite{Bacchetta:2006un}
\begin{equation} 
A_{UT}^{\sin(\phi_{R} + \phi_S)\sin \theta} (x, Q^2) =
- C_y\, 
\frac{\sum_q e_q^2\,h_1^q(x, Q^2)\, n_q^{\uparrow} (Q^2) }
        {\sum_q e_q^2\,f_1^q(x, Q^2) \, n_q (Q^2)} \; , 
\label{eq:asydis}
\end{equation} 
where we consider only the $x$ (the momentum fraction of the initial quark) binning as our interest here lies  on the transversity distribution.
The depolarization factor is 
$C_y  
                   \approx
\frac{1-\langle y \rangle}
       {1-\langle y \rangle +\langle y \rangle^2/2}$ ;
and
\begin{align} 
n_q (Q^2) = \int  dz dM_h^2\,D_1
(z, M_h^2, Q^2) \, ; \quad 
n_q^{\uparrow} (Q^2) = \int  dz \, dM_h^2
\frac{|\bf{R}|}{M_h} \,H_{1, sp}
(z, M_h^2, Q^2) \, , 
\label{eq:npair}
\end{align}
with $|{\bf R}| / M_h =\sqrt{1/4 - m_\pi^2 / M_h^2}$ . 
$D_1$ is the unpolarized DiFF describing the hadronization of a quark $q$ 
into a $\pi^+\pi^-$ pair plus any number of undetected hadrons, averaged over
quark polarization and pair orientation. Finally, $H_{1, sp}^{\sphericalangle}$ is a chiral-odd 
DiFF, and denotes the component of $H_1^{\sphericalangle }$ that is sensitive to the 
interference between the fragmentation amplitudes into pion pairs 
in relative $s$ wave and in relative $p$ wave. 

Isospin symmetry and charge conjugation allow for the assumptions~\cite{Bacchetta:2006un}:
\begin{gather}
D_{1}^{u} = D_{1}^{d} = D_{1}^{\bar{u}} = D_{1}^{\bar{d}} \; , \quad
D_1^s = D_1^{\bar{s}} \; , \quad D_1^c = D_1^{\bar{c}} \; , \nonumber
\\
H_{1, sp}^{\sphericalangle u} = - H_{1, sp}^{\sphericalangle d} = - H_{1, sp}^{\sphericalangle \bar{u}} = 
H_{1, sp}^{\sphericalangle \bar{d}} \; , \label{eq:ass3}
\end{gather}
the rest being $0$.
We also assume $D_1^s  \equiv N_s\, D_1^u$ and we consider the two scenarios 
$N_s = 1$ and $N_s = 1/2$. The second choice is suggested 
by the output of the PYTHIA event generator. 
The difference between these 2 scenarios defines the theoretical error of our result.

The above assumptions allow us to turn Eq.~\eqref{eq:asydis} into the
following simple relation (neglecting charm quarks) 
\begin{equation} 
x h_1^{u_v}(x, Q^2) - {\textstyle \frac{1}{4}}\, x h_1^{d_v}(x, Q^2) = 
- \frac{A_{\text{DIS}}(x, Q^2)}{C_y} 
\,  \frac{ n_u (Q^2) }{n_u^{\uparrow} (Q^2)} 
\sum_{q=u,d,s} \frac{e_q^2 N_q}{e_u^2} x f_1^{q+\bar{q}}(x, Q^2) \; , 
\label{eq:simple}
\end{equation} 
where $N_u = N_d = 1$ and $f_1^{q+\bar{q}} = f_1^q + f_1^{\bar{q}}$,  
$h_1^{q_v} = h_1^q - h_1^{\bar{q}}$. 
Our goal is to derive from data the difference between the 
valence up and down transversity distributions by computing the r.h.s. of the above
relation. 
 
The unpolarized PDFs in Eq.~(\ref{eq:simple}) can be estimated using any parametrization 
of the unpolarized distributions. We chose to employ the MSTW08LO PDF 
set~\cite{Martin:2009iq}.

The only other unknown term on the r.h.s.\  of Eq.~(\ref{eq:simple}) is the ratio 
$n_u / n_u^{\uparrow}$. We extract this information from the recent measurement 
by the Belle collaboration of the Artru--Collins azimuthal asymmetry 
$A^{\cos (\phi_R + \bar{\phi}_R)}$~\cite{Vossen:2011fk}.
            %
%
For this purpose we need $n_u^{\uparrow} / n_u$ at the experimental values of 
$\langle Q^2 \rangle$ 
and integrated over the HERMES 
invariant-mass range $0.5 \leq M_h \leq 1$ GeV. We will get to this number in two steps. 

{\it First}, 
 the ratio $n_u^{\uparrow} / n_u$ is integrated over $0.5 \leq M_h \leq 1$ GeV at 
 the Belle scale (100 GeV$^2$). 
The relevant variables for DiFFs in electron-positron annihilation are $(z, M_h, \bar z, \bar M_h)$. Belle's data are differential in different pairs of variables. We consider the Belle asymmetry integrated over $(z, \bar{z})$ and binned in 
$(M_h, \bar{M}_h)$. We weight each relevant bin by the unpolarized cross section, 
which, in this process, is approximated by  the inverse of the statistical error squared. 
By summing over all bins in the considered range, we get the total asymmetry 
\begin{align}
 A^{\cos (\phi_R + \bar{\phi}_R)}&= \frac{-\langle \sin^2 \theta_2 \rangle}{\langle 1+\cos^2 \theta_2 \rangle} \, 
\frac{ \langle \sin \theta \rangle \, \langle \sin \bar{\theta} \rangle \, 5 \, (n_u^{\uparrow})^2}
        { (5+N_s^2)\, n_u^{2} + 4\, n_c^{2} }  
= - 0.0307 \pm 0.0011 \; ,
\label{eq:asye+e-int}
\end{align} 
which leads to,~\footnote{
Useful values extracted from Belle analysis are given in Ref.~\cite{Bacchetta:2011ip}.}
\begin{equation}
n_u^{\uparrow} / n_u (100\, \mathrm{GeV}^2) = - 0.273 \pm 0.007_{\mathrm{ex}} \pm 
0.009_{\mathrm{th}} \; , 
\label{eq:ratioQ0}
\end{equation} 
where the second error comes from using the two different values of the $s-$quark normalization 
$N_s$.~\footnote{
To verify the reliability of this procedure, we repeated the calculation 
estimating the denominator of the asymmetry using the PYTHIA
event generator
without acceptance cuts.}

{\it Second}, since the Belle scale is very different from the HERMES one, 
DiFFs must be connected from one scale to the other via their QCD evolution 
equations~\cite{Ceccopieri:2007ip}. In order to do this, we need to know the $z$ 
dependence of  $H_{1, sp}^{\sphericalangle\, u}$ and $D_1^q$ for each $M_h$ value.  We start from a parametrization of both DiFFs at $Q_0^2 = 1$ GeV$^2$, evolving at LO using the HOPPET code~\cite{Salam:2008qg} and fitting Belle's data.  The fitting procedure will be further explained elsewhere.~\footnote{ We checked that the final 
results are affected in a negligible way by the gluonic component $D_1^g (z,M_h;Q_0^2)$. }
By integrating the extracted DiFFs in the HERMES range $0.5 \le M_h \le 1$ GeV and 
$0.2 \le z \le 1$, we can calculate the evolution effects on $n_u^{\uparrow}/n_u$ at each 
$\langle Q^2 \rangle$. 
It turns out that the ratio is 
decreased by a factor $0.92 \pm 0.08$, where the error takes into account the 
difference of $Q^2$ in the HERMES experimental bins as well as the uncertainty related to 
different starting parametrizations at $Q_0^2 = 1$ GeV$^2$. In conclusion, for the extraction of transversity in Eq.~(\ref{eq:simple}) we use 
the number
\begin{equation}
n_u^{\uparrow} / n_u = - 0.251 \pm 0.006_{\mathrm{ex}} \pm 0.023_{\mathrm{th}} \; .
\label{e:ratioQ}
\end{equation} 
%
%
\begin{figure}
  \includegraphics[height=.25\textheight]{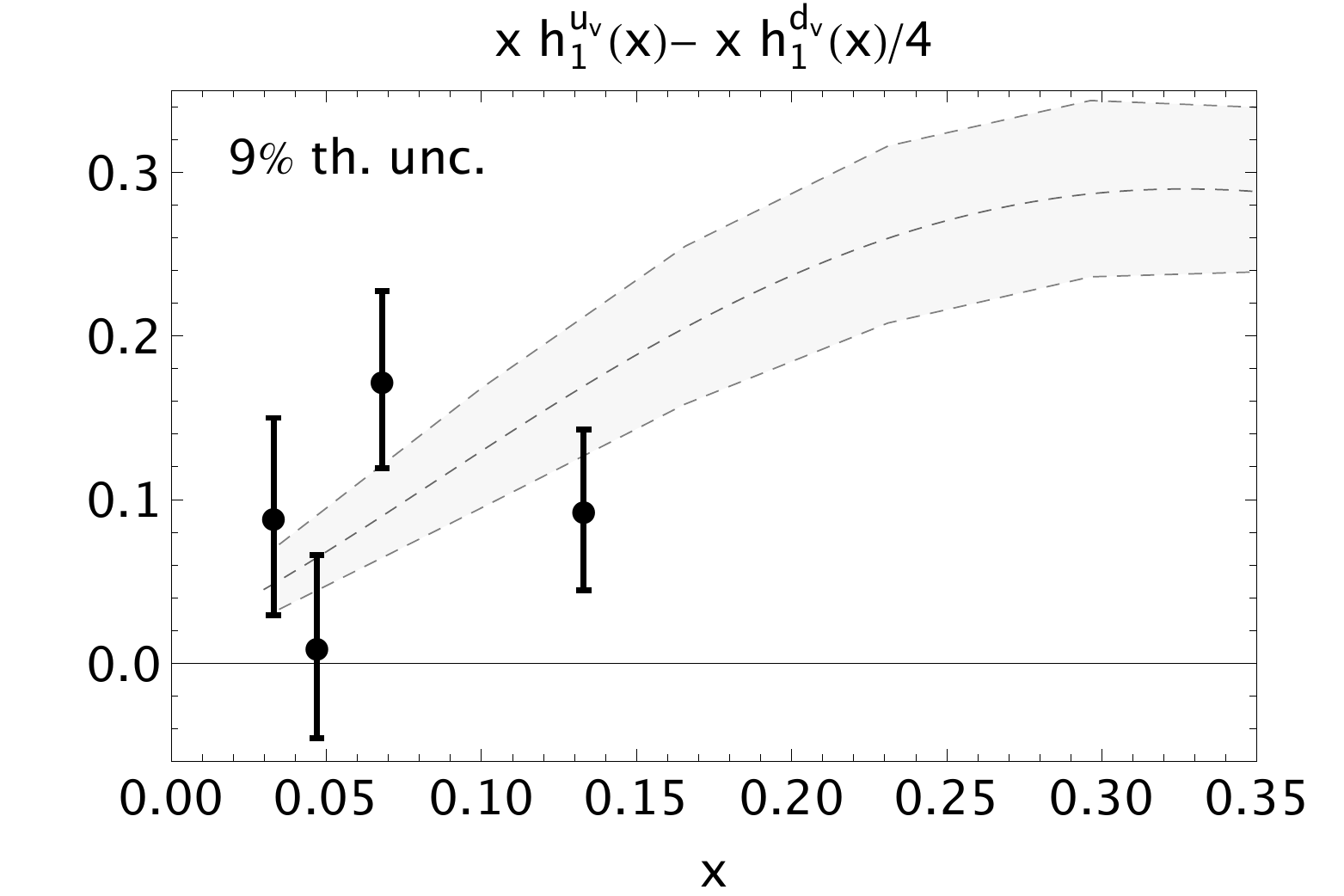}
  \caption{\label{fig:h1param} The $x h_1^{u_v} - x h_1^{d_v} / 4$ of Eq.~(\ref{eq:simple}) as a 
function of $x$. 
}
\end{figure}
Our result for the four HERMES data points is shown in 
 Fig.~\ref{fig:h1param}. 
%
The central line represents the best fit for the combination 
$x h_1^{u_v} - x h_1^{d_v} / 4$, as  deduced from the most recent parametrization of 
$h_1^{u_v}$ and $h_1^{d_v}$ extracted from the Collins effect~\cite{Anselmino:2008jk}. 

In summary, we have presented a determination of the transversity parton 
distribution in the framework of collinear factorization by using data for pion-pair 
production in deep inelastic scattering off transversely polarized targets, combined with data 
of $e^+ e^-$ annihilations into pion pairs. The final trend of the extracted transversity 
seems not to be in disagreement with the transversity extracted from the Collins 
effect~\cite{Anselmino:2008jk}. More data are needed to clarify the issue. 






\bibliographystyle{aipproc}   

\bibliography{mybiblio}

\IfFileExists{\jobname.bbl}{}
 {\typeout{}
  \typeout{******************************************}
  \typeout{** Please run "bibtex \jobname" to optain}
  \typeout{** the bibliography and then re-run LaTeX}
  \typeout{** twice to fix the references!}
  \typeout{******************************************}
  \typeout{}
 }

\end{document}